\documentclass[12pt]{article}

\usepackage{amsmath, setspace}
\usepackage{amssymb, xspace}

\usepackage{graphicx}
\usepackage{color} 

\usepackage{subcaption}
\usepackage{hyperref}
\usepackage{times}
\usepackage[numbers, sort&compress]{natbib} 
\usepackage[table,rgb]{xcolor}
\usepackage{multirow}
\usepackage[misc,geometry]{ifsym}

\usepackage[font=footnotesize]{caption}
\DeclareCaptionLabelSeparator{pipe}{$\; | \;$}
\definecolor{abstarctBlue}{rgb}{0.0706, 0.349, 0.6667} 
\definecolor{abstarctBlue2}{cmyk}{ 0.5143,   0.2857,    0.0286,    0.3714}
\definecolor{Section}{cmyk}{0.2,0.8,0.8,0.3}

\usepackage{soul}
\sodef\an{\fontfamily{phv}\selectfont}{.08em}{1em plus1em}{0.5em plus.1em minus.1em} 
\sodef\ann{\fontfamily{phv}\selectfont}{0.04em}{0.5em plus0.02em}{0.1em plus.1em minus.1em}

\makeatletter
\renewcommand{\@biblabel}[1]{\quad#1.}
\makeatother

\date{}

\usepackage{overpic}
\newcommand*{\hvfont}{\fontfamily{phv}\selectfont}

\newcommand{\fg}{\textcolor{linkcolor}{Fig.}~\ref}

\newcommand{\pana}{({\bf a})\xspace}
\newcommand{\panb}{({\bf b})\xspace}

\definecolor{citecolor}{rgb}{0.071, 0.36, 0.67}   
\definecolor{linkcolor}{rgb}{0.071, 0.4, 0.67}  
\hypersetup{
colorlinks=true, 
citecolor=citecolor,
linkcolor=linkcolor,
urlcolor=linkcolor
}

\let\citep=\cite
\let\citet=\cite

\begin{document}
\include{Captions}
\begin{spacing}{1.55}
\noindent {\LARGE \bf 
Analyzing ribosome remodeling in health and disease 
}
\end{spacing}

\noindent\ann{
Aleksandra A.~Petelski,$^{1,2,3}$
\& Nikolai Slavov$^{1,2,3,}$\textsuperscript{\Letter }
} \\

{\small 
\noindent 
$^{1}$Department of Bioengineering, Northeastern University, Boston, MA 02115, USA\\
$^{2}$Barnett Institute, Northeastern University, Boston, MA 02115, USA\\
$^{3}$Department of Biology, Northeastern University, Boston, MA 02115, USA\\
}
{\scriptsize \Letter} Correspondence: \href{mailto:nslavov@alum.mit.edu}{\an{\small nslavov@alum.mit.edu}} or \href{mailto:nslavov@northeastern.edu}{\an{\small nslavov@northeastern.edu}} \\
\thispagestyle{empty}

\vspace{8mm}
\noindent{\bf
Increasing evidence suggests that ribosomes actively regulate protein synthesis. However, much of this evidence is indirect, leaving this layer of gene regulation largely unexplored, in part due to methodological limitations. Indeed, we review evidence demonstrating that commonly used methods, such as transcriptomics, are inadequate because the variability in mRNAs coding for ribosomal proteins (RP) does not necessarily correspond to RP variability. Thus protein remodeling of ribosomes should be investigated by methods that allow direct quantification of RPs, ideally of isolated ribosomes. We review such methods, focusing on mass spectrometry and emphasizing method-specific biases and approaches to control these biases. We argue that using multiple complementary methods can help reduce the danger of interpreting reproducible systematic biases as evidence for ribosome remodeling. 
}
\vspace{8mm}

\begin{spacing}{1.3}
\section*{Introduction}
The control of gene expression is crucial for all biological processes, from developmental stages and homeostasis maintenance to regeneration processes. This regulation occurs at multiple layers, both transcriptional and post-transcriptional levels.  Historically, transcription has been studied more extensively than translation, in large part because of the accessibility of technologies for nucleic acid analysis. However, gene regulation via translation was appreciated as early as the late 1960's. For example, the production of insulin was linked to the increased number of polysomes and protein synthesis \citep{wool1968mode}. Similarly, the synthesis of globin (hemoglobin polypeptides) was found to depend on heme, the oxygen-carrying iron-rich molecule in the blood \citep{levere1965control}. In the early 1980s, different aspects of early embryonic development, such as blastocyst formation \citep{braude1979control} and disparate patterns of protein synthesis \citep{rosenthal1980selective} were linked to post-transcriptional regulation. Key developmental genes, such as Oscar, are regulated at the level of translation and spatial localization \citep{markussen1995translational, martin2009mrna}. Such post-transcriptional regulation is mediated at least in part by translational control mechanisms including RNA binding proteins, translation factors, and micro RNAs \citep{castello2012insights, bartel2009micrornas}.   

In addition to these well-established mechanisms, increasing evidence suggests that ribosomes may also regulate translation of mRNAs \citep{mauro2002ribosome_filter, komili2007functional, gilbert2011functional, Slavov2015-em,preiss_2015, Emmott_Slavov_Ribo_TIBS_2019}. Ribosomes have long been viewed as passive players in translation, with a fundamental role of catalyzing peptide-bond formation but exerting regulatory effects only based on their availability \citep{lodish1974model, mills2017ribosomopathies} and translational control being exerted by a teamwork of cis- and trans-regulators in which the interaction of mRNA structures and sequences worked with translation factors, RNA binding proteins, and microRNAs; the ribosome complex was simply viewed as an effector of translation. However, ribosome-mediated regulation through the alteration of ribosomal RNA and ribosomal proteins were both hypothesized \citep{naora1999involvement, mauro2002ribosome_filter, gilbert2011functional} and supported by mostly indirect evidence as recently reviewed by Emmott et al. \citep{Emmott_Slavov_Ribo_TIBS_2019}. Quantification of ribosomal proteins from isolated ribosomes has begun to provide more direct evidence, as mammalian ribosomes have been found to exhibit differential protein stoichiometry that depends on the growth conditions and on the number of ribosomes per mRNA\citep{Slavov_ribo}. Such observations suggest that different ribosome complexes may exist in order to fulfill disparate functions, which can consequentially have regulatory effects on translation. This model, termed ribosome specialization, challenges the notion that ribosomes are static enzymes and instead introduces them as active participants in post-transcriptional regulation. In this model, ribosomes could be specific to distinct cell population and can affect the translation of mRNAs. The hypothesis of ribosome-mediated translational regulation has been further supported by observations of differences in ribosomal protein composition occurring under stress \citep{Ferretti2017-uo, Samir2018-of, ghulam2020differential} and cell differentiation\citep{kawasaki2011nitration,akanuma2012inactivation, khajuria2018ribosome, Shi2017-cz}.  Localized synthesized RPs in the axon can also contribute to ribosome remodeling \citep{Cagnetta2018-rl, Shigeoka2019-ja}, suggesting possible roles of ribosome-mediated translational control in neuronal functions. The possibility of ribosome specialization is further supported by the observations that mutations of specific ribosomal proteins selectively affect the synthesis of specific proteins and are strongly associated with distinct phenotypes such as cancer and aging \citep{steffen2008RP_aging, lawrence2014discovery}. Such selectivity suggests that ribosomes can regulate gene expression \citep{komili2007functional, xue2012specialized, gilbert2011functional, Emmott_Slavov_Ribo_TIBS_2019}. Interestingly, some RPs, such as RACK1, have been observed to dynamically  associate and dissociate from ribosomes and specifically affect the translation of short mRNAs \citep{gerbasi2004yeast, thompson2016ribosomal, johnson2019rack1}.

Specialized ribosomes have been suggested to hold specific functional roles, especially in the context of immunology and cancer. The idea of  ribosomes driving cancer progression could be tied to observations of disease states characterized by dysfunctional ribosomes, disorders collectively known as ribosomopathies \citep{draptchinskaia1999gene, liu2006ribosomes}. Patients with such disorders showed increased risk for diseases of uncontrolled cell growth, such as cancer, later in life \citep{de2015ribosomopathies}. More direct evidence has shown the association of RP gene mutations with numerous cancers, raising the prospect of the existence of oncoribosomes \citep{sulima2017ribosomes, kampen2018rise}. Ribosomes have been also implicated in the immunosurveillance of cancer and other types of pathogenic cells. MHC class I molecules, which are important in alerting the immune system when cells are virally infected, are believed to be derived from DRiPs (defective ribosomal products), unstable molecules that degrade much more quickly than functional proteins that are in stable conformation \citep{yewdell1996defective}. In order to rapidly produce such products, a subset of ribosomes could be assigned to synthesize DRiPS that have enhanced antigen presentation \citep{yewdell2006immunoribosomes, yewdell2019immunoribosomes}. Immunoribosomes may also serve as an efficient source of peptides that can stimulate antibody production upon the invasion of pathogenic molecules. These models of functionally specialized immunoribosomes and oncoribosomes remain insufficiently tested, and further testing would benefit from the approaches discussed in this review.

Ribosomes in eukaryotes are made of about 80 core ribosomal proteins and four ribosomal RNAs (rRNAs) \citep{warner1999economics}. The modifications of both molecular groups are important in the functionality of the ribosome complex and are potential sources of ribosome heterogeneity and specialization \citep{sloan2017tuning, penzo2016importance}. Modifications of rRNA, such as methylation and pseudouridylation, which together span a total of ~7000 nucleotides, are known to stabilize the ribosome structure. Variant rRNA alleles and rRNA methylation may contribute to ribosomal specialization. In both mice and humans, several rRNA sequence variants were identified and shown to exhibit tissue-specific expression; furthermore, at least 23 percent of rRNA nucleotides are estimated to exhibit variant alleles within the general human population \citep{parks2018variant}. On average, 32 variants were found to be expressed in single individuals, while those sequence variants were found to significantly overlap with sequences that are functionally important to ribosome function. The diversity of rRNA variants are thus suggested to to have a biologically important role. Indeed, the methylation of the 16S rRNA at a specific guanosine nucleotide revealed that this modification plays a role in controlling mistranslation and could explain streptomycin-resistant phenotypes of {\it M. tuberculosis} \citep{wong2013functional}. Both rRNA and protein modifications may contribute to ribosome specialization; The role of rRNA has been reviewed by Mauro and Matsuda \citep{mauro2016translation}, and in this review we will focus on methods for investigating ribosome specialization via modifications of the core ribosomal proteins. 

 \begin{figure}[h!]
   \centering
   \begin{overpic}[width = .66\textwidth]{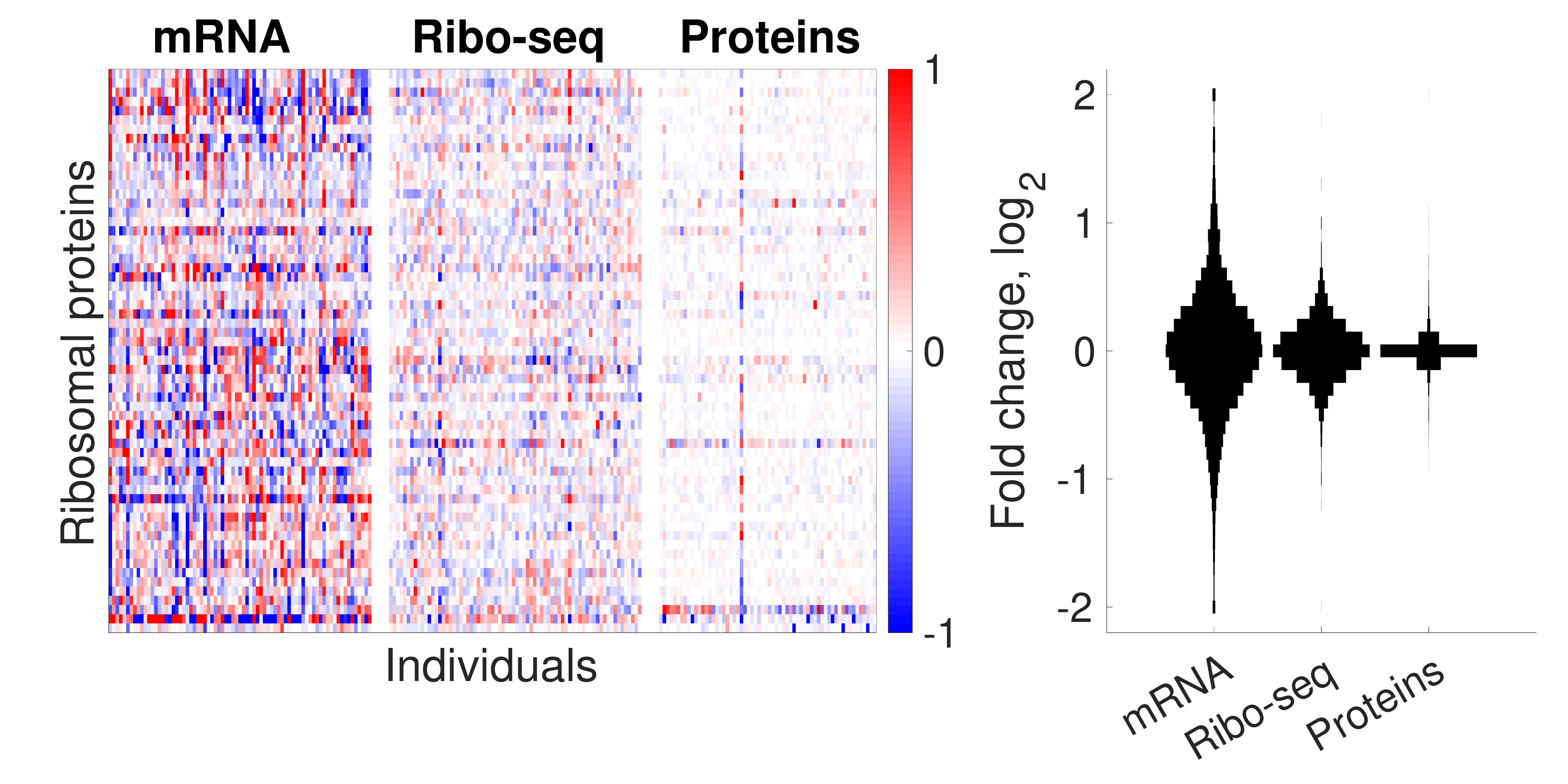} 
          \put(-2,46){\large \bf \hvfont a}
      \end{overpic} 
      \hspace{.04\textwidth}
      \begin{overpic}[width = .28\textwidth]{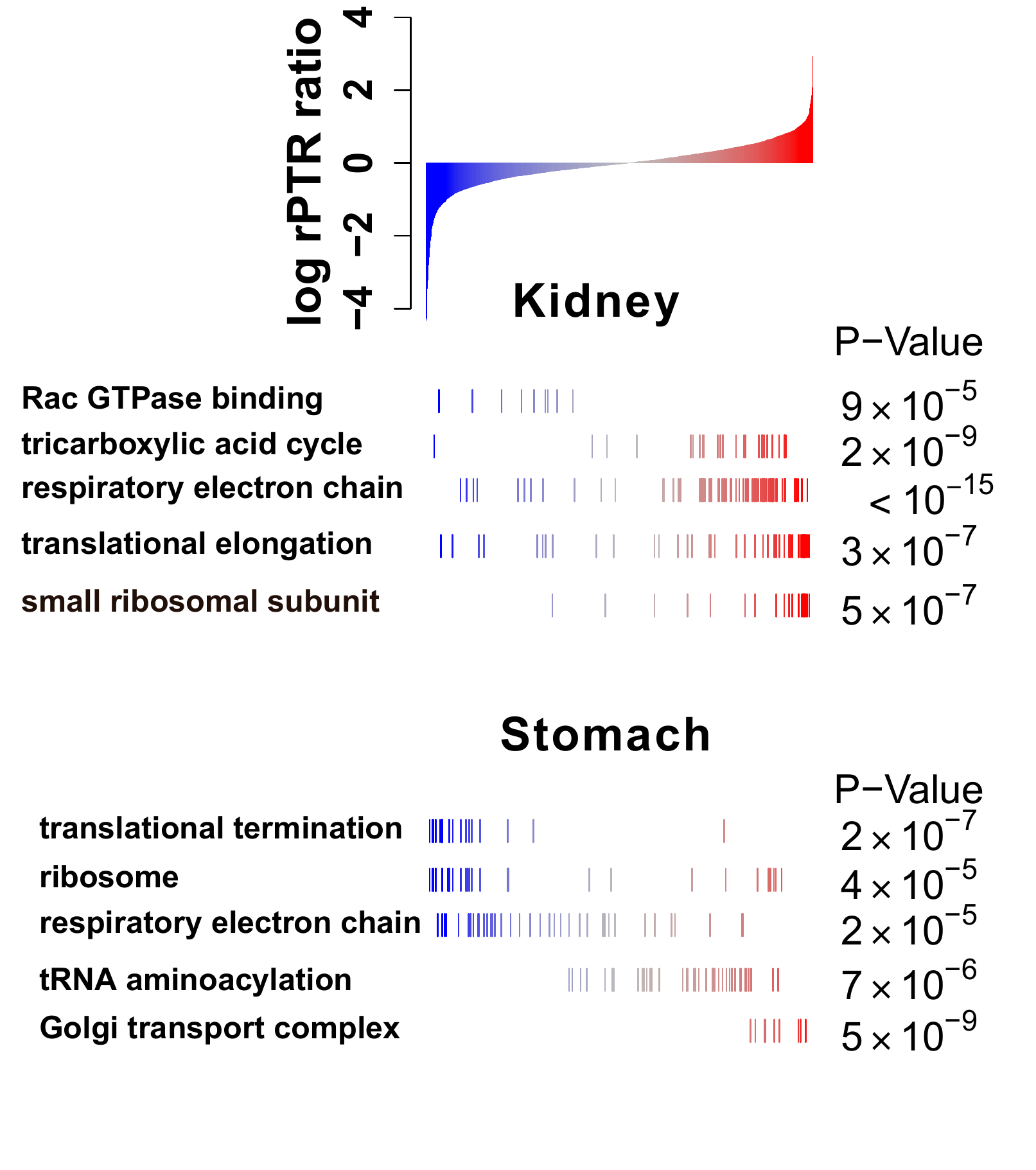} 
          \put(-1,100){\large \bf \hvfont b}
      \end{overpic} 
   \caption{ \hvfont {\bf Ribosomal proteins are under post-transcriptional control}
   \pana A heatmap showing the levels of ribosomal genes at the levels of RNA, ribosome density, and protein in lymphablastoid cell lines based on data from Battle {\it et al.} \citep{Battle2015-mb}. The data points for each gene are displayed as $log_2$ fold-changes relative to their mean.
   The corresponding distributions of fold changes for all genes across all cell lines are shown to the right. To control for variable input amounts from different cell lines, the data from each cell line were normalized to the same total amount of RP gene products.   
   \panb Protein to mRNA ratios (PTRs) were quantified across different human tissues, and gene sets with statistically significant shifts in the tissue-type specific relative protein to RNA ratios (rPTR) are highlighted. Of interest, ribosomal proteins have high rPTRs in the kidney and low rPTRs  the stomach, indicating significant post-transcriptional regulation. This panel is based on the analysis by Franks {\it et al.} \citep{Franks2016PTR}. 
   }
   \label{RP_PTR}
 \end{figure}

\section*{Post-transcriptional regulation of ribosomal proteins}
While suggestive evidence for ribosome remodeling has originated from indirect methods (such as measuring transcripts coding for RPs \citep{RPL38_kondrashov2011ribosome, Slavov_eth_grr}), such data remain inconclusive because RP synthesis and degradation are extensively regulated \citep{sung2016ribosomal, Eisenberg2018-kp,Franks2016PTR, Battle2015-mb}. RP molecules that are not incorporated into a complex are rapidly degraded\citep{sung2016ribosomal}. Because of this post-transcriptional regulation, analysis of ribosome remodeling in health and disease should rely on direct protein measurements. While the abundance of RP transcripts are usually the most accessible data, these measures are indirect. Consider, for example, the variability of RP transcripts, ribosome density, and ribosomal proteins across a panel of lymphoblastoid cell lines shown in \fg{RP_PTR}a. The data indicate substantial transcript variability, which diminishes at the level of ribosome density and is almost absent for the ribosomal proteins, \fg{RP_PTR}a. Thus, variable RP transcripts do not necessarily indicate variable abundance of RPs, \fg{RP_PTR}a. More generally, the ratio between RPs and their corresponding transcripts can vary substantially as shown in \fg{RP_PTR}b for two tissues and observed in many other cases \citep{zhang2014proteogenomic, Franks2016PTR}.  Therefore, the levels of mRNAs coding for RPs are rather indirect evidence to support conclusions about the protein composition of ribosomes.

To obtain more direct evidence for protein remodeling of ribosomes, one should directly measure RP abundances as a more direct approach for evaluating ribosome remodeling across different conditions, \fg{RP_PTR}a.  Such analyses in cell lysates have suggested  changes in the protein composition of ribosomes as budding yeast undergoes the diauxic shift \citep{Slavov_exp}, during aging \citep{janssens2015protein,Kelmer_Sacramento2020-us}, and upon LPS-stimulation of mouse dendritic cells\citep{jovanovic2015dynamic}. In addition, RP analysis from cell lysates has shown 
 that ribosomal transcripts exhibit slower elongation rates with decreased protein production relative to other transcripts with similar ribosome densities\citep{riba2019protein}. 

Although RPs are known to degrade very quickly when not incorporated into complexes\citep{sung2016ribosomal}, RPs quantified in total cell lysates may originate in part from  other extraribosomal complexes \citep{naora1999involvement, warner2009common, statello2018identification, weisberg2008transcription, bhavsar2010other, zhou2015ribosomal}. Excluding the influence of such extraribosomal complexes requires the analysis of isolated ribosomes. The use of isolation methods, such as sucrose gradient fractionation or affinity purification\citep{dougherty2017expanding}, prior to protein measurements provides an even more direct way to assess ribosome remodeling in different conditions \citep{shigeoka2019site, Slavov_ribo, hummel2012dynamic}.      

In some studies, the measurement noise is comparable to the variability of RPs across conditions \citep{amirbeigiarab2019invariable}. In such cases, we may conclude that ribosomes do not remodel across the examined conditions or the degree of remodeling it too small to be detected by the methodology used. The smaller the measurement errors, the more confident we may be that ribosomes do not remodel across the set of studied conditions.
For example, the degree of ribosome remodeling (if any at all) is very small during the  aging of mouse brains \citep{amirbeigiarab2019invariable}. As observed RP changes can be small, they may be comparable to or smaller than the measurement noise.

\begin{figure}[h!]
\centering
      \begin{overpic}[width = .6\textwidth]{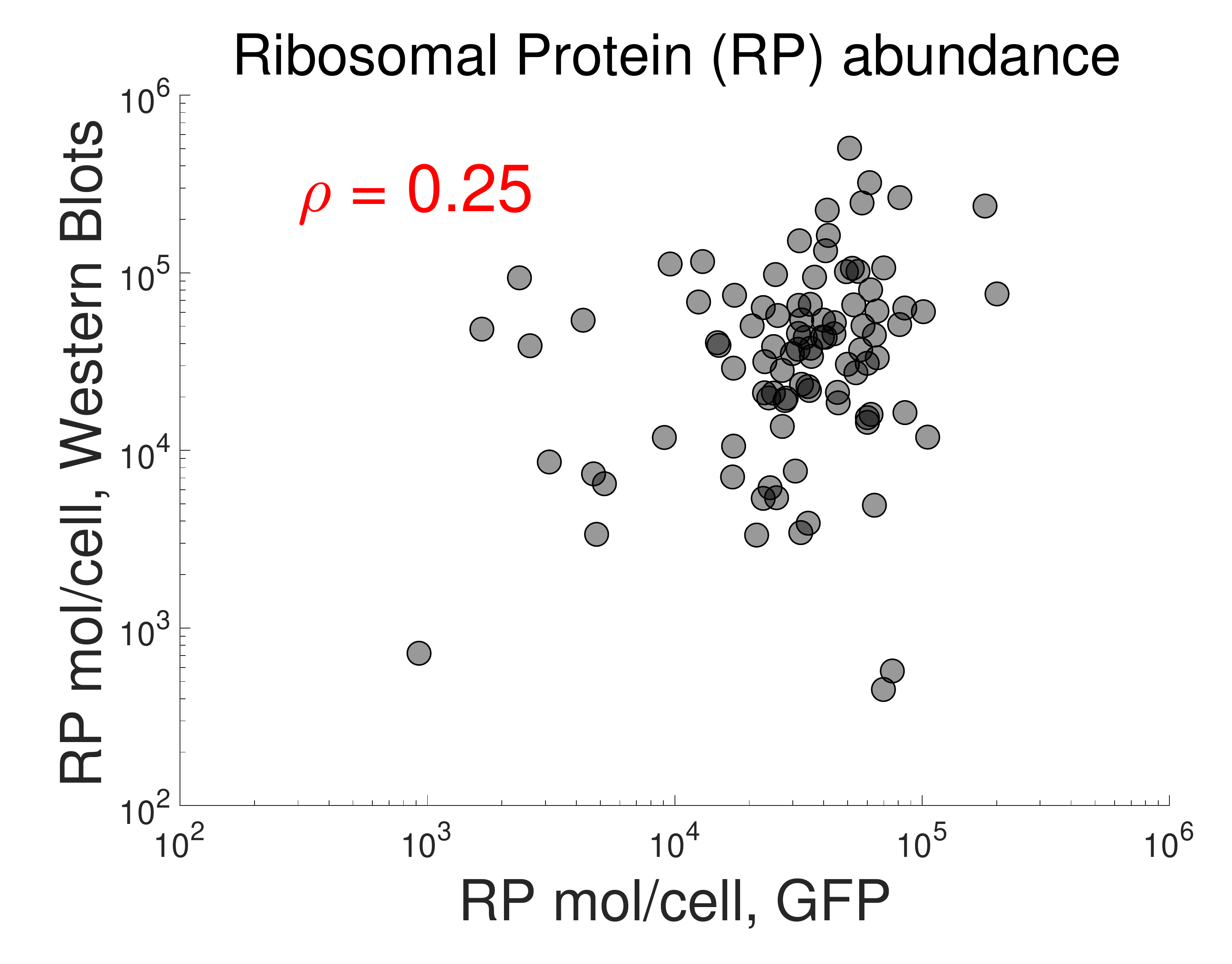} 
      \end{overpic} 
	\caption{\hvfont {\bf Different methods provide different estimates of ribosomal protein abundance.}  
	 A scatter plot comparing estimated abundance of ribosomal proteins from Western blots \citep{Ghaemmaghami2003-rf} and from flow cytometry analysis of RPs tagged by green fluorescent protein (GFP)\citep{Newman2006-zr}. The Pearson correlation between the two estimates is modest, $\rho = 0.25$, despite the fact that the measurements by each method are highly reproducible. The weak correlation many reflect shared biases from the construction of the tagged proteins or biological differences in protein abundance. The first option seem much more likely, but it remains inconclusive without additional data and analysis.   
	 }
	\label{GFP_Western}
\end{figure}

\section*{Precise and reproducible measurements may not be accurate}
Since each method comes with its own potential for biases and systematic artifacts, the characterization of ribosome remodeling also calls for the use of complementary methods. A highly precise method can offer consistently reproducible measurements that are also consistently biased. Reproducibility does not necessarily correspond to accuracy. For example, RP levels estimated by GFP-tagging and by Western blots differ significantly (\fg{GFP_Western}a) despite the fact that replicates within each method are reproducible \citep{Ghaemmaghami2003-rf, Newman2006-zr}. Generally, measurements can be affected by systematic biases, leading to technically reproducible but inaccurate measurements. Such data may consistently support an incorrect representation of the studied biological systems. The effect of biases is especially important to recognize when studying ribosome remodeling, since  the changes of ribosomal protein stoichiometry are often relatively small as observed in previous studies \citep{Slavov_ribo,Shi2017-cz, ferretti2017rps26}.

\begin{figure}[h!]
\centering
      \begin{overpic}[width = .85\textwidth]{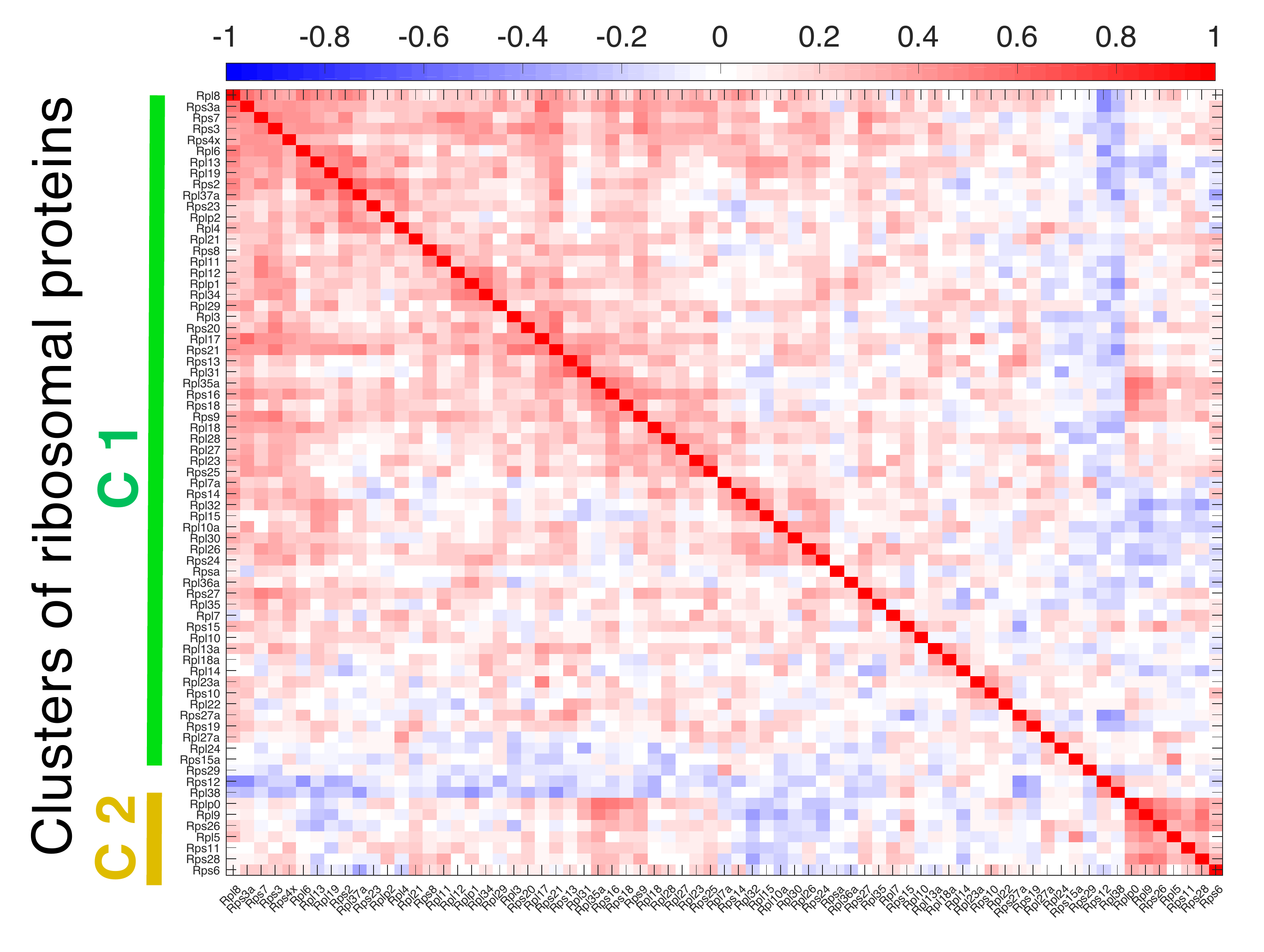} 
      \end{overpic} 
	\caption{\hvfont {\bf Covariation of ribosomal proteins across diverse single cells}  
	Pairwise Pearson correlations between RPs computed from protein levels measured by SCoPE-MS in individual differentiating stem cells \citep{scopems2017, scopems2018}. The correlation matrix was clustered based on the cosine between the correlation vectors. This panel is reproduced from Budnik {\it et al.} \citep{scopems2017}.
	  }
	\label{single_cell_RPs}
\end{figure}

\section*{Analyzing ribosome remolding in individual cells}
Ribosome remodeling is likely to contribute to the specialized proteomes of the diverse cell types that arise during development and diverge to form different tissues \citep{Emmott_Slavov_Ribo_TIBS_2019}. Furthermore, cyclic transcription of mRNAs coding for RPs suggests that ribosome biogenesis is highly temporarily organized and metabolically coordinated during the cellular life cycle \citep{Slavov_batch_ymc, Slavov_pop, Slavov_emc}. 
While these possibilities are of considerable interest, their investigation poses particular challenges for direct RP quantification by complementary methods  \citep{Slavov2020Science,slavov2020Review}. These challenges stem from the difficulty of quantifying proteins in tissues comprised of heterogeneous cells. 
Single-cell proteomics has approached this question in the context of embryonic stem cells differentiating to epiblast lineages in embryoid bodies \citep{scopems2018}. In this system, the correlations among RPs (\fg{single_cell_RPs}) revealed one large and one small cluster, suggesting that the RPs from the small cluster covary in a cell-type specific manner. However, these data have not been cross-validated by independent methods and thus remain inconclusive. Nonetheless, advances in single-cell proteomics hold the potential to increase the reliability of single-cell protein analysis and to enable cross-validation of such measurements \citep{specht_high-throughput_2019, Specht_Perspective_2018, Slavov2020Science,slavov2020Review}. Thus these technological advances may soon enable direct examination of ribosome specialization within the diverse cell types that comprise different tissues.

\section*{The need for complementary methods}
The technical biases of each method type are challenging to overcome, but the use of complementary methods with divergent weaknesses can help guard against the potential influence of systematic artifacts on the final results. Thus, the use of complementary techniques allows for both the attainment of strong evidence for ribosome remodeling and the cross-validation of results that can support more confident conclusions. As an example, RPs from isolated ribosomes may be quantified separately by both mass spectrometry and Western blots and then the results can be compared as shown in \fg{MS_WB}. The measured RP differences obtained from these two disparate methods are highly similar, indicating that the results are likely to be driven by biological effects rather than the inherent biases attributed to each method. More broadly, if biases were the main force behind the results, the outcomes would have been presumably different; thus, the more different the methods, the less likely they are to share biases, and the more beneficial they are in being used together as complementary methods.   



\newcommand{\IIms}[2]{ \begin{overpic}[width = .27\textwidth]{#1} \put(-4,105){\Large \bf \hvfont  #2}\end{overpic} }
\newcommand{\IIwb}[2]{ \begin{overpic}[width = .6\textwidth]{#1} \put(0,70){\Large \bf \hvfont #2}\end{overpic} }

\begin{figure}[ht!]
 \begin{tabular}{>{\centering\arraybackslash} p{.38\textwidth} %
 				 >{\centering\arraybackslash} p{.62\textwidth}}
 	  \Large Mass--Spec  &  \Large Western Blots \\ [1.5em]
 	\IIms{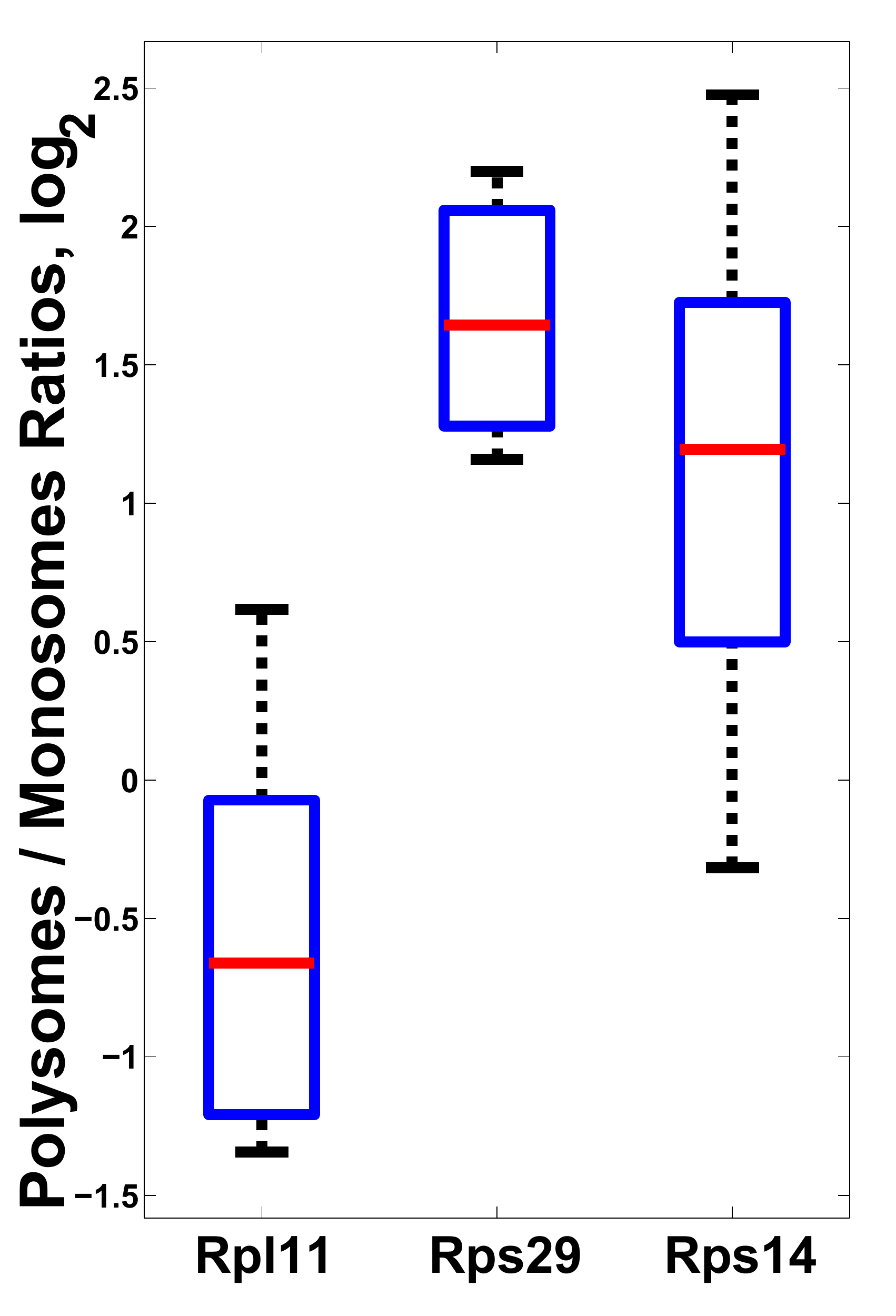}{a} &
 	\IIwb{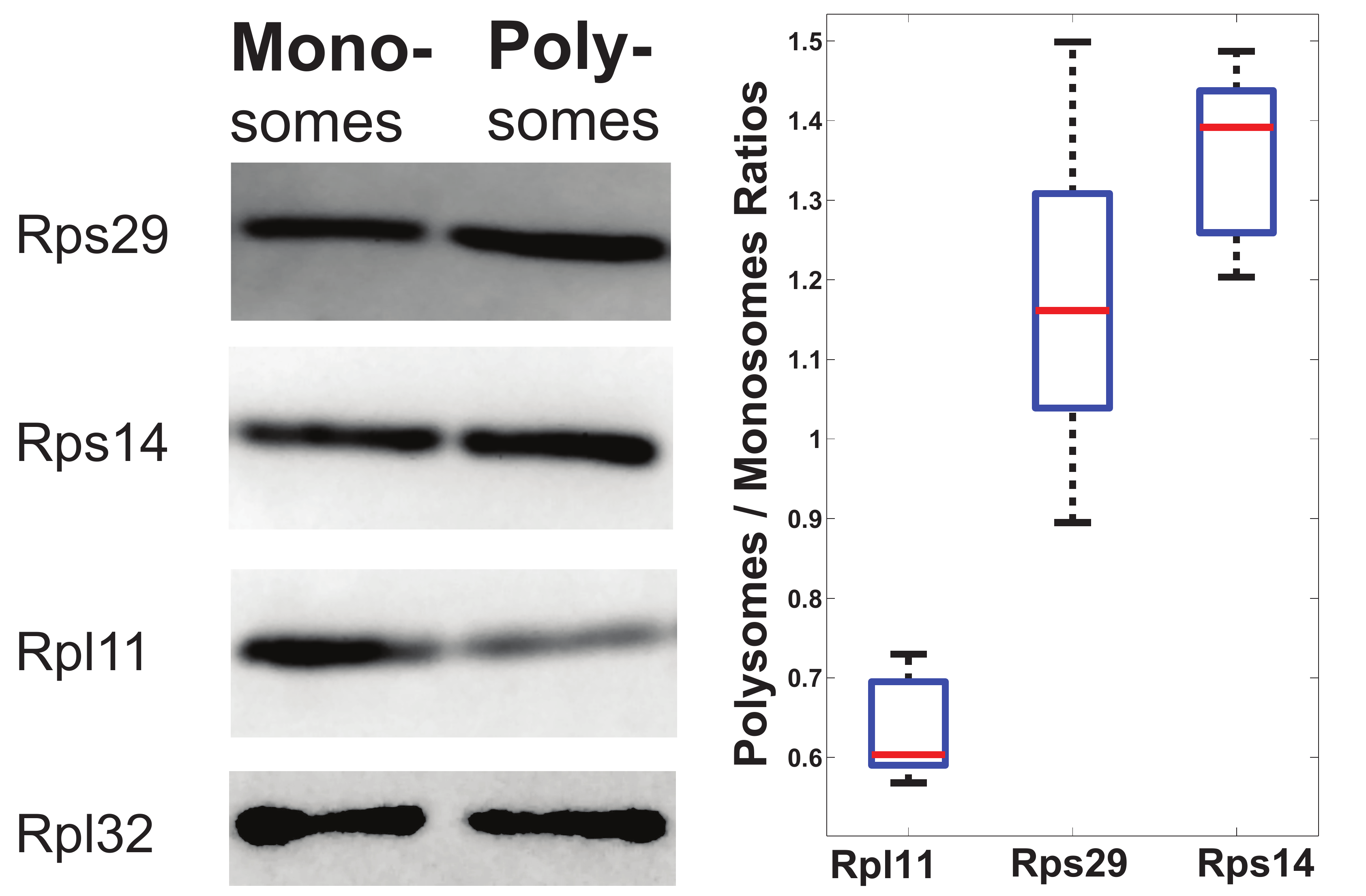}{b} \\
  \end{tabular}	
   \caption{ {\bf Quantifying ribosomal proteins with complementary methods} Monosomal and polysomal ribosomes from embryonic stem cells were isolated by sucrose gradients and analyzed by mass-spectrometry and by Western blots \citep{Slavov_ribo}. The results exemplify qualitative agreement for the polysomal enrichment of Rps29 and Rps14 as well as relatively small differences across the two methods. The differences may be due both to the different cell lines used or to the quantification biases of each method  \citep{Slavov_ribo}.   
 \pana The Polysomal enrichment of RPs was quantified by mass-spectrometry using isobaric tags.  The analysis was performed by digesting the RPs either with trypsin or with Lys-C and the results provided consistent estimates that are combined in these boxplots \citep{Slavov_ribo}.  
\panb Polysomal enrichment of RPs quantified by Western blots \citep{Slavov_ribo}. RPs were quantified by Western blots in monosomes and polysomes from E14 mouse ESCs. Rpl32 was used as a loading control and the boxplots summarize data from 9 ratios for each quantified RP. The panels are reproduced from Slavov {\it et al.} \citep{Slavov_ribo}.  
}
\label{MS_WB}
\end{figure}

Direct quantification of RPs with complementary methods (as shown in \fg{MS_WB}) can reveal changes in the ribosome complex that may contribute to post-transcriptional regulation of gene expression \citep{Slavov_ribo}. Mass spectrometry (MS) offers an array of such complementary approaches for discovering and validating ribosome remodeling. These powerful methods directly measure protein or peptide abundance, either through relative or absolute quantification, and can provide information about the stoichiometry of the ribosome complex along with modifications of the ribosomal proteins. These results can be enhanced and cross-validated by structural methods, such as  X-ray crystallography and cryogenic electron microscopy (cryo-EM). These latter methods can directly probe the arrangement of ribosomal proteins within the complex. In this review, we discuss the types of biases, limitations, and advantages associated with each approach type, focusing particularly on the investigation of ribosome remodeling.

\section*{Quantifying ribosomal proteins via mass spectrometry}
The levels of ribosomal protein expression can be directly measured using mass spectrometry proteomics. Whole proteins, or even entire ribosome complexes can be analyzed by top-down mass spectrometry \citep{catherman2014top, toby2016progress}. However, because MS analysis of full length proteins is technologically challenging, it is more common to first digest proteins into peptides and then analyze the peptides by MS. This latter approach is known as bottom-up mass spectrometry, which involves the caveat of inferring  protein levels from digested peptides. These two branches (top-down and bottom-up) of mass spectrometry can be implemented by many methodologies, each of which has systematic biases and can directly  quantify RP stoichiometry across biological conditions. 

Top-down and bottom-up proteomics methods provide complementary measurements, as they quantify proteins in inherently different ways and thus can help detect and mitigate biases. Technical biases that can affect the ultimate biological interpretation of any experiment can be potentially introduced at any step of the proteomics pipeline, from cell lysis and sample preparation to MS analysis. These distinct biases can be controlled by using different methods. Ultimately reliable results require cross-validating the results from different methods that share as few biases as possible. 


\subsection*{Biases in bottom-up mass spectrometry}
Bottom-up proteomics can be implemented by many methods, all of which have caveats, specifically for quantifying  ribosomal proteins. The amount of ribosomes in most cells is generally ample, about 10 million in a typical mammalian cell. Yet, the sequences of ribosomal proteins are rather  short, with a median length of about 133 amino acids in human cells, in stark contrast to the overall proteome sequence length, which is 375 amino acids. Thus, ribosomal proteins tend to produce fewer peptides during digestion. In addition, ribosomal proteins are likely to produce shorter peptides when using trypsin, the most commonly used protease, due to a large number of arginine (R) and lysine (K) amino acids present in the protein sequences \citep{lott2013comparative}. The over representation of K and R may contribute to a higher miscleavage rate, which can complicate analyses. These downsides can be taken into account by the use of multiple alternative proteases such as Lys-C and Glu-C in order to cross-validate results with the commonly used protease of trypsin \citep{giansanti2016six,Multi-Enzymatic-Limited-Digestion}.  These biases can be very significant when estimating absolute protein abundances (due to differences in peptide flyability, a collective term describing efficiency of ionization and detection), and these biases may cancel out in relative protein quantification \citep{peng2012protease, HIquant-MCP}.

\subsection*{Quantifying protein stoichiometry by bottom-up mass spectrometry}
The differences in peptide flyability and other peptide specific biases poise a major challenge to quantifying stoichiometry (ratios) between different proteins and their proteoforms \citep{peng2012protease, Giansanti2015-zv}. Indeed, a peptide might be more intense because it is delivered more efficiently to the MS analyzer rather than because it originates from a more abundant protein. These peptide specific biases cancel out when performing relative quantification of a protein across different samples. This idea of canceling out biases can be extended by first-principle models, such as HIquant, to allow quantifying stoichiometry between different proteins independent from peptide specific biases \citep{HIquant-MCP}. Such an approach is likely to be particularly fruitful for quantifying RP preforms originating from different prologues, alternative splicing events or post-translational modifications \citep{Emmott_Slavov_Ribo_TIBS_2019,emmott2019approaches,HIquant-MCP}.

In bottom-up proteomics, cells can be lysed and proteins extracted by a variety of methods \citep{hughes2014ultrasensitive, mPOP_2018, sielaff2017evaluation}. Then proteins are  digested to peptides, which are separated via liquid chromatography (LC) and ionized through electrospray ionization (ESI) or matrix-assisted desorption/ionization (MALDI). Then, these peptides are introduced into the mass spectrometer as precursor ions, which can be used for direct quantification or can be further isolated and disintegrated into fragment ions that are then used for sequence identification and quantification. Bottom-up MS may either simultaneously isolate and fragment multiple peptides in parallel (known as DIA; data independent acquisition), or analyze a single peptide at a time (known as DDA; data dependent acquisition). DDA methods were introduced in 1990s and have been widely used for decades \citep{cravatt2007biological, Aebersold_Mann_2016}. DIA was introduced later (2004) by Yates and colleagues \citep{Venable2004-mf} and has matured into methods that offer the advantage of parallel analysis and afford identifying and quantifying many thousands of peptides in a single run \citep{Demichev2020-fv,Muntel2019-zm}.  

Quantification of peptides can be based directly on precursor ions (MS1-based) or fragment ions (MS2-based). 
Both of these peptide quantification approaches can be implemented by multiple methods. These methods differ in the way and extent to which they control technical biases stemming from the proteomics pipeline, from sample preparation procedures to LC-MS analysis. Wet-lab procedures encompass the experimental steps of isolating ribosomes from cell lysates, protein digestion into peptides, and various labeling techniques. In LC-MS analysis, biases can be introduced due to differential peptide separation, ionization, and isolation for MS2 analysis. The available bottom-up methods control for these technical biases at different stages. Due to the disparities in biases, different bottom-up approaches can be viewed as highly complementary techniques that allow us to gain more confidence in results showing biological changes, especially in ribosome remodeling where the changes may be small. 


\subsubsection*{Control of biases at the MS1-Level of Peptide Quantification }
Methods that use the MS1 level in order to quantify peptides can be further classified either as labeled or label-free approaches. The label-free approaches are attractive for their simplicity and fewer experimental steps. In these methods, each sample undergoes the sample preparation and mass spectrometry analysis separately. In contrast, labeled techniques allow analyzing multiple samples in parallel and thus more opportunities to control for biases that occur during the parallel stages.  Labeling approaches for MS1-based quantification include: SILAC (Stable Isotope Labeling of Amino Acids in Cell Culture), dimethyl labeling, and mTRAQ. SILAC introduces sample labeling to living cells during protein synthesis through the metabolic labeling of newly synthesized proteins\citep{ong2002stable}. More specifically, cells are incubated in cell culture medium that contains stable-isotope enriched amino acids, usually arginine and lysine \citep{blagoev2004temporal, berger2002high}, which are then incorporated into new proteins. Dimethyl labeling, on the other hand, introduces sample labeling after digestion. This chemical labeling approach relies on the reaction between formaldehyde and sodium cyanoborohydride with lysine side chains and N-terminal primary amines in order to form dimethylamines \citep{hsu2003stable, Boersema2009-aw}. mTRAQ quantification, which also introduces labels after digestion. It was designed specifically for targeted MS, in which specific peptides found in previous shotgun runs could be probed \citep{desouza2009absolute, kang2010quantitative}. All of these methods quantify peptides at the level of the precursor ions.

The introduction of biases can be controlled through mixing the samples as early as possible within the experimental pipeline. The earlier the samples are combined together, the sooner the samples can be exposed to identical conditions and the sooner biases can be controlled. Among the available MS methods, SILAC allows mixing of samples earliest, even prior to ribosome isolation and MS procedures, thus enabling the control of all biases associated with these experimental and MS steps. This is especially advantageous when studying subcellular fractions, such as the ribosome complex. Labeling methods, such as dimethyl labeling and mTRAQ, allows the mixing of samples prior to LC separation and MS analysis, which controls for LC- and MS-related biases, while label-free methods will not control for these biases, as each sample is analyzed individually.

The processes of peptide separation via liquid chromatography and subsequent MS analysis can introduce technical biases that  MS1-based labeling methods can control but label-free cannot. Each run in the mass spectrometer is subject to a host of variable factors that might influence measured ion intensities, including variability in peptide separation, ionization, and instrumentation.  This can cause different runs to experience drifts in retention time and m/z, leading to complicated analyses. Ion suppression, a problematic phenomenon that affects the final amount of charged ions that ultimately reaches the detector, can also undermine quantitative accuracy. In the cases of labeling, digested peptides from each sample are separated, ionized, and analyzed together, allowing for the samples to experience the same nuances associated with each of those steps. Label-free approaches in particular suffer from the fact that large portions of peptides are not detected in every sample, which is termed as the missing value problem\citep{webb2015review}. This problem makes it more difficult to compare RP abundance across different conditions. Advances relying on matching peptide intensity readouts, i.e. MaxLFQ\citep{cox2014accurate}, and on enhanced peptide identifications via DIA  \citep{Venable2004-mf,Demichev2020-fv,Muntel2019-zm}  or  DDA methods incorporating retention time information \citep{dartID_PLoS, yu2020isobaric, slavov2020Review} can mitigate the missing value problem when using label-free approaches. 

While MS-1 based labeling methods allow controlling for biases during peptide separation and ionization, labeling itself can introduce biases, limitations, and artifacts. SILAC introduces heavy isotopes into live cells and animals, which can induce unintended growth changes\citep{crotty2011differential, filiou201215n} and may even affect behavioral characteristics of mice\citep{frank2009stable, filiou201215}. Also, depending on the model system used, the potential conversion of arginine to proline may lead to underestimation of heavy-labeled peptides\citep{van2007experimental}. An additional limitation when using SILAC is that the time to fully label a biological system of interest can take days to weeks. Dimethyl labeling, on the other hand, takes around 5 minutes and is highly cost-effective, as the reagents are inexpensive \citep{boersema2009multiplex}. However, dimethyl labeling has been associated with a loss of hydrophillic peptides\citep{lau2014comparing}, which can lead to fewer overall peptide identifications. Overall, any type of labeling approach should be assessed for completeness of labeling to allow for confident quantitation of RP remodeling. 


MS1-based quantification techniques share some common limitations.  A disadvantage of MS1-based labeling methods is that the number of multiplexed samples is usually limited to 2 or 3. As the number of isotopically labeled samples increases, so does the number of precursor ions in MS1 scans. The high density of ions may lead to interference between coeluting ions with very similar m/z ratios.  The use of NeuCode metabolic labeling offers higher SILAC multiplicity through the use of labels that differ in mass on the scale of milli-Daltons\citep{merrill2014neucode, overmyer2018multiplexed}; thus, this technique requires higher resolving power of the mass analyzers \citep{merrill2014neucode, overmyer2018multiplexed}.  Because of the inherent limitations that MS1-level methods present, it is important to consider using complementary methods rather than multiple technical replicates of just one method type. MS1-based methods can be co-validated with MS2-based approaches, which still come with their own biases; however, the set of biases derived from both quantification approaches are different, allowing for increased confidence in quantifying even small RP changes.

\subsubsection*{Control of biases at the MS2-Level of Peptide Quantification}
As opposed to MS-1 based methods, precursor ions are isolated and then  fragmented, and the resulting (fragment) ions analyzed by another (MS2) scan. In most cases, MS2-based quantification is used with isobaric tags, such as tandem mass tags (TMT)\citep{thompson2003tandem} or Isobaric Tags for Relative and Absolute Quantification (iTRAQ)\citep{ross2004multiplexed, choe20078}. DIA analysis can also benefit from incorporating MS2-level data in order to improve the accuracy of label-free quantification \citep{Huang2020-fk}. 

With isobaric labeling, peptides of each sample are covalently labeled with a sample-specific mass tag. Mass tags are comprised of a reporter ion, a balance group and a reactive group, which is usually an amine reactive group (NHS group) that binds to primary amines from the peptides.   After labeling, the samples are mixed together, allowing the analysis  multiple samples (up to 16 with TMT Pro) in a single mass spectrometry run. Thus, the multiplexed samples undergo ionization and ion selection for MS2 analysis together, greatly mitigating the missing value problem.  Additionally, the multiplexing permits analyzing more samples per unit time. Another advantage is that the number of precursor ions detected during MS1 scans does not increase with the number of samples, as each isobaric tag has an identical mass which allows a particular peptide from different samples to appear as a single feature in the ion map defined by retention times and  m/z ratios. Peptides from different samples are thus indistinguishable at the MS1 level. Upon isolation and fragmentation of precursor ions, they release distinct reporter ions and peptide fragments, some of which remain bound to the balance group.  Since the reporter ions and the balance groups have sample-specific number of heavy isotopes, they allow for the ratiometric quantification of samples.

Since all peptide from a sample release the same reporter ion upon fragmentation, the accuracy of quantifying a peptide from its reporter ions is dependent on isolating only its precursor ion for subsequent fragmentation and MS2 analysis. In practice, the MS2 scan typically also contains coisolated peptides. The degree of coisolation can be estimated and used to remove peptides with unacceptably high coisolation. Narrower isolation windows \citep{specht_high-throughput_2019} and sampling elution peaks at their apexes \citep{doMS2019} reduce coisolation but may not completely eliminate it. When multiple peptide precursors are coisolated for fragmentation, the measured reporter ion intensities are proportional to the  superposition of peptide abundances, which results in inaccurate quantification and generally compressed peptide ratios.  However, ratio compression is not necessarily due to coisolation since it may be caused by other factors, such as unintentional cross-labeling or sample carry-over on the LC column. Coislation can be reduced by subjecting fragment ions to further isolation and fragmentation steps by taking MS3 or even higher scans\citep{ting2011ms3}. This MS3 approach reduces the number of analyzed ions and thus diminishes the sensitivity and the throughput of the analysis. A second option for reducing biased due to coisolation is to use the peptide fragments with attached balance groups, which are termed  complement ions\citep{sonnett2018accurate}. These complement ions are produced during the fragmentation step as a result of the mass balance group remaining attached to the peptide or its fragments, and thus they can be specific to the analyzed peptide.  

Abundant peptides, such as those originating from RPs, tend to be less affected by coisolation since the majority of the reporter ions used for quantification will be derived from the abundant peptides. However, coisolation can still contribute significant bias to RP quantification and thus necessitates quality controls. One way to benchmark data quality is through the calculation of the agreement of different peptides originating from the same protein. Ribosomal peptides that originate from a RP should indicate consistent quantification of the protein; the degree of consistency  can be quantified by measures such as reliability or coefficients of variation \citep{Franks2016PTR}.   

The methods of quantifying peptides at either the MS1 or MS2 levels come with inherent biases that might complicate analyses and may present reproducible results that are actually artifacts induced from systematic biases. However, the set of biases for each method type is different, and thus different methods may complement each other. Instead of choosing just one mass spectrometry method to identify changes of ribosomal proteins, methods that differ as much as possible (such as MS1 and MS2 based quantification) or non mass-spec methods should be used in parallel (as shown in \fg{MS_WB}) so that the results can be co-validated.

\subsection*{Top-down mass spectrometry}
With top-down proteomics\citep{catherman2014top, toby2016progress}, whole ribosomal proteins or even whole ribosomes can be analyzed, offering an more intact picture of the ribosome and potential characterization of proteoforms and post-translational modifications (PTMs). These methods identify the full amino acid sequence of a protein including modifications and thus do not require protein inference from peptides. However, these techniques are more more challenging because compared to peptides, proteins are more difficult to solubilize, separate, ionize and sequence \citep{chait2006mass}. These challenges result in lower sensitivity and throughput of the analysis, as well as higher technical requirements from the instruments for high resolving power at high m/z ratios \citep{van2006improving, Van_de_Waterbeemd2017-vi, heck2008native}. Nonetheless, these challenges are rapidly being addressed by innovative methods for protein separation (such as capillary zone electrophoresis \citep{Belov2017-dc}), sensitive methods allowing the detection of individual ions \citep{Neil2020SingleIons,worner2020resolving}, and by community standards \citep{Donnelly2019-gb}.   

The biases associated with either bottom-up or top-down proteomics are very different, which presents opportunities for combining them synergistically. The hybrid combination of mass spectrometry techniques has been used to characterize whole ribosome complexes by quantifying RP levels along with proteoforms \citep{Van_de_Waterbeemd2017-vi}. Such an approach has offered novel observations about  cysteine modification of RPS27, and ribosome assembly sites through the characterization of ribosomes originating from human, plant, and bacterial cells \citep{van2018dissecting}. Additionally, the elusive ribosomal protein SRA was successfully quantified and found to have heterogeneous stoichiometry in {\it E. coli} ribosomes \citep{Van_de_Waterbeemd2017-vi}. 


\section*{Complementary structural biology techniques}
We have focused our review on the higher throughput MS methods, but many structural biology methods can provide more detailed information about ribosome modifications and structure albeit at lower throughput. These include chemical cross linking of proteins \citep{sinz2018cross, yu2018cross}, X-ray diffraction, and cryo-EM. 
The analysis of both protein sequence in tandem with spatial information allows not only to co-validate results ribosome remodeling observations but also to start revealing its functional significance. Such a combined approach has been used to study other complexes, such as the human nuclear pore complex scaffold \citep{npc_2013_beck}, bacteriophage portal complexes\citep{Poliakov2007Feb}, 26S proteasome complex\citep{Lasker2012Jan} and COPII vesicles that transport proteins to the Golgi apparatus from the endoplasmic reticulum \citep{noble2013pseudoatomic}. In fact, such combination of structural biology and mass spectrometry has been used to study a particular subcomplex involved in ribosome biogenesis \citep{davis2016modular, wu2017atomic}. The synergy between these methods can help identify specialized ribosomal structures and confirmations regulating protein synthesis.  


X-ray diffraction requires the crystallization of the sample of interest prior to analysis, while cryo-EM eliminates this need. The process of crystallization can demand much optimization, which can involve the purification of many ribosomes. The obviation of crystallization allows for the study of more kinds of ribosomes without the limitation of crystallizing them. The flash-freezing of samples allows the observation of molecules in a "near-native" state; since molecules are not constrained to a crystal, they have more degrees of freedom, thus allowing for more conformational states to be studied. However, cryo-EM studies can require anywhere from hundreds to thousands single particle images, which involves long hours of image acquisition at the microscope. Furthermore, many of these images are discarded due to the phenomenon known as beam-induced motion, which produces blurred images \citep{nogales2015cryo}. Despite limitations, both technique types continue to reveal the ribosome protein structure. Recently, X-ray diffraction was used to discover the role of potassium ions within the ribosome on the stabilization of the protein complex in both the initiation and elongation states of translation \citep{rozov2019importance}. Cyro-EM has been nearing the sensitivity and resolution of X-ray diffraction, as seen through the report of the entire bacterial ribosome resolved at two angstroms \citep{watson2020structure}. 


The synergy of mass spectrometry and structural biology techniques can empower novel observations of the ribosome complexes. These method types are different in terms of sample preparation and measurement acquisition, yet are complementary in the type of information that is provided. Combining protein composition information with conformational changes can help reveal different roles of ribosomes within protein synthesis, and ultimately gene expression. Such knowledge can help decipher the degree of ribosome remodeling during normal  development and physiology (e.g., immunoribosomes) and during diseases (e.g., oncoribosomes). The combination of mass spectrometry with structural biology has been used to elucidate several facets of the ribosome complex subunits, including ribosome assembly in bacteria\citep{mulder2010visualizing, sashital2014combined, davis2016modular} and the effect of dimerization on ribosomes when nutrients are scarce \citep{feaga2020ribosome}. In addition, ribosome remodeling has been found to occur in the bacterial ribosome through the use of MS and X-ray crystallography \citep{lilleorg2019bacterial}.  

This paper emphasized the need to use multiple complementary approaches for quantifying ribosome remodeling and briefly reviewed sources of systematic biases as well as approaches for mitigating their influence. These approaches may afford reliable quantification of ribosome remodeling, which is a starting point for investigating its functional roles in regulating mRNA translation. Identifying ribosomes associated with specific conditions -- such as disease states, developmental stages, or metabolic conditions -- can start to reveal different populations of ribosomes. These observations will serve as a starting basis for characterizing new principles in the regulation of RNA translation that may reshape our understanding of one of the most fundamental biological processes. In the long term, this new understanding can enable the design of therapies that specifically target translation for disease and regenerative treatments.


\end{spacing}

\bigskip
\noindent {\bf Acknowledgments:} This work was funded by a New Innovator Award from the NIGMS from the National Institutes of Health to N.S. under Award Number DP2GM123497.\\ 

\bibliographystyle{plos-natbib} 
\bibliography{SCoPE,ribo,Review,mPOP}

\begin{thebibliography}{148}
\bibitem[{Sinz(2018)}]{sinz2018cross}
Sinz A (2018) Cross-linking/mass spectrometry for studying protein structures
  and protein--protein interactions: where are we now and where should we go
  from here?
\newblock Angewandte Chemie International Edition 57: 6390--6396.
\end{thebibliography}

\end{document}